# Leaky Mode Engineering: A General Design Principle for Dielectric Optical Antenna Solar Absorbers

Yiling Yu[2], Linyou Cao[1,2]*

[1]Department of Materials Science and Engineering, North Carolina State University, Raleigh NC 27695; [2]Department of Physics, North Carolina State University, Raleigh NC 27695;

**Abstract**

We present a general principle for the rational design of dielectric optical antennas with optimal solar absorption enhancement: leaky mode engineering. This builds upon our previous study that demonstrates the solar absorption of a material with a given volume only dependent on the density and the radiative loss of leaky modes of the material. Here we systematically examine the correlation among the modal properties (density and radiative loss) of leaky modes, physical features, and solar absorption of dielectric antenna structures. Our analysis clearly points out the general guidelineS for the design of dielectric optical antennas with optimal solar absorption enhancement: a) using 0D structures; b) the shape does not matter much; c) heterostructuring with non-absorbing materials is a promising strategy; d) the design of a large-scale nanostructure array can use the solar absorption of single nanostructures as a reasonable reference.

The strong, tunable optical resonance of subwavelength dielectric structures presents a promising strategy for the enhancement of solar absorption in semiconductor materials.[1-6] The dielectric structure can act as an optical antenna to efficiently trap incident solar light into a confined space, which can subsequently lead to strong solar absorption in semiconductor materials with a reduced volume. Unlike the surface plasmon resonance in metallic structures,[7-11] in which the absorbed solar energy may be lost as heat owing to the lossy nature of metals, dielectric optical antennas can make sure that the incident solar radiation can be absorbed only by semiconductor materials and be converted to extractable charge carriers. Recent research has demonstrated substantial optical resonances and solar absorption enhancements with a variety of dielectric structures. These include nanoparticles, nanowires, nanopillars, nanocones, and nanowells.[5, 12-41]

Despite the extensive studies on dielectric optical antenna solar absorbers, there are no principles established that can guide the rational design of the antenna to achieve optimal solar absorption enhancement. It is well recognized that the solar absorption efficiency strongly depends on the physical features, including dimensionality, morphology, structure, and composition, of the antenna.[5, 12-31] To achieve the optimal solar absorption enhancement in a material would request surveying and comparing the solar absorption over all the possible physical features of the material. However, the existing designs rely on rigorous numerical or analytical models, such as Mie theory, rigorously coupled wave analysis (RCWA), finite difference time domain (FDTD), to find out the amplitude of solar absorption as a function of physical features. These methods involve intensive computation efforts and show limited capabilities of designing dielectric optical antennas with increasing structural complexity. They are only able to optimize the solar absorption in a relatively narrow parameter window. The lack of general design principles that would allow for the optimization of solar absorption over a broad range of physical features has delayed the exploitation of the full potential of dielectric optical antennas for the development of cost-effective solar cells.

Here we present a general principle for the rational design of dielectric optical antenna solar absorbers: leaky mode engineering. Leaky modes are defined as natural optical modes with propagating waves outside the antenna structure. Each mode is featured with a complex eigenvalue ($N_{real} - N_{imag}.i$) that can be solved analytically or numerically.[42-44] This study builds upon our previous work that demonstrates the solar absorption of a material is solely dictated by the modal properties (density and radiative loss) of leaky modes in the material.[45] In this work, we systematically elucidate the correlation between the modal properties of leaky modes and the physical features (dimensionality, shape, etc.) of dielectric antenna structures. We also demonstrate that the difference in the solar absorption of various dielectric structures can be ascribed to the difference in the modal properties of leaky modes. This work points out that keys to maximize the solar absorption in a material are to increase the density of leaky modes and to tune the radiative loss of the leaky modes to reasonably match the intrinsic absorption loss of the material.

We start with briefly reviewing the key discoveries of our previous studies.[45] We have demonstrated that the light absorption in semiconductor structures of any dimensionality (zero, one, and two) can be evaluated using an intuitive model, coupled leaky mode theory (CLMT).[42] The CLMT model considers the light absorption as a result from the coupling between incident light and the structure's leaky modes (Figure 1a inset). The solar absorption of a single-mode semiconductor structure with known refractive index is found only dependent on two variables of the mode, the radiative loss $q_{rad}'$ and the resonant wavelength $\lambda_0$, regardless the physical features of the structure. The two variables are related with the complex eigenvalue ($N_{real} - N_{imag}.i$) of the leaky mode as $q_{rad}' = N_{imag} / N_{real}$ and $\lambda_0 = 2\pi nr / N_{real}$, where $n$ and $r$ are the refractive index and characteristic size of the structure, respectively. As an example, Fig. 1a shows the calculated solar absorption of a single-mode 1D structure as functions of the radiative loss $q_{rad}'$ and the resonant wavelength $\lambda_0$. Amorphous silicon (a-Si) is used as the absorbing material in this structure.[46] The solar absorption is calculated by integrating the spectral absorption cross-section $C_{abs}(\lambda)$ of the single-mode structure over the spectral photon flux of solar radiation $I(\lambda)$ as $P_{solar} = \int C_{abs}(\lambda) I(\lambda) d\lambda$. For the convenience of discussion, we assume that every absorbed photon generates one electron and the 1D dimensionality dictates the calculated solar absorption in a unit of $mA/cm$.

The solar absorption of a multi-mode structure, as what are typically used in solar cells, is a simple sum of the absorption contributed by each of the modes, $P_{total} = \int P_{solar}(\lambda_0) \rho(\lambda_0) d\lambda_0$, where $P_{solar}(\lambda_0)$ and $\rho(\lambda_0)$ are the solar absorption of one leaky mode and the density of leaky modes at an arbitrary resonant wavelength $\lambda_0$, respectively. To maximize the solar absorption requests the absorption of each mode to be optimized. We can illustrate this notion using the result of 1D a-Si structures given in Fig. 1a. From Fig. 1a, we can identify the optimal absorption of single leaky modes with an arbitrary resonant wavelength as indicated by the white dashed line. We replot the single-mode optimal absorption and associated radiative loss as a function of the resonant wavelength $\lambda_0$ in Fig.1b. The maximum solar absorption of a multi-mode a-Si 1D structure can be obtained by performing integration over this single-mode optimal absorption as illustrated by the shaded area shown in Fig.1b.

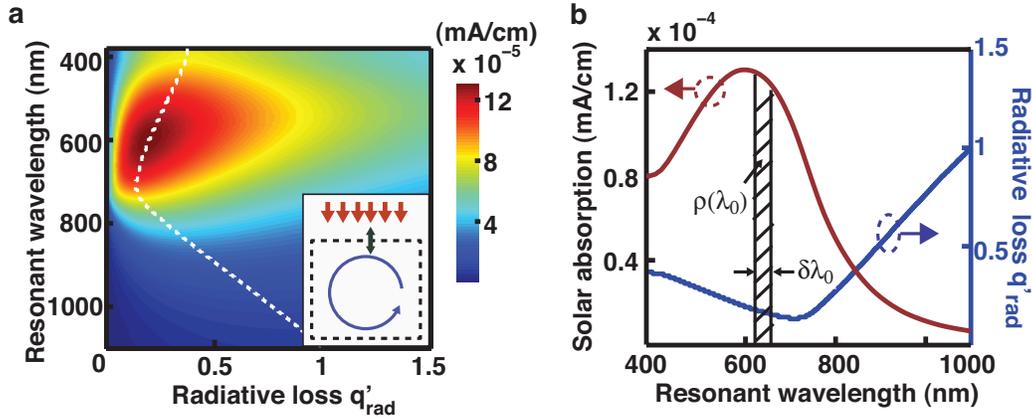

Figure1. (a) Calculated solar absorption of single-mode 1D structures as a function of the radiative loss and resonant wavelength of the mode. a-Si is used as the absorbing material in the 1D structure. The white dashed line connects the optimal absorption at each resonant wavelength. The inset schematically illustrates the coupling of incident light with the leaky mode in nanostructures. (b) The single-mode optimal absorption (red line) and associated radiative loss (blue line) as a function of the resonant wavelength. The shaded area schematically illustrates the integration of contributions from multiple leaky modes.

The deterministic correlation of solar absorption with the leaky modes suggests general guidelines for the rational design of dielectric optical antenna solar absorbers. As the total solar absorption is a sum of the contribution from each leaky mode, the density of leaky modes $\rho(\lambda_0)$ poses a fundamental limit for the solar absorption. Additionally, we can find in Fig. 1 that the absorption of single leaky modes is particularly strong when the resonant wavelength and radiative loss of the leaky mode are tuned into the ranges of 400 - 800 nm and 0.05-0.7, respectively. Intuitively, the range of 400 - 800 nm for the resonant wavelength is to match the spectral photon flux of solar radiation, and can generally apply to all kinds of active materials for solar absorption, including a-Si, CdTe, CIGS, and organic materials. The range of 0.05-0.7 for the radiative loss is to reasonably match the intrinsic absorption loss (defined as $n_{imag}/n_{real}$, $n_{real}$,

$n_{imag}$ are the real and imaginary part of the refractive index, respectively) of the active materials, a-Si. The match of the radiative loss and the absorption loss can create a desired "critical coupling" between incident solar radiation and the materials.[47] For a material with a given volume, the number of leaky modes in the desired spectral range (400-800nm) is determined by the density of the modes. Therefore, keys to maximize the solar absorption in a material are to increase the density of leaky modes and to tune the radiative loss of the leaky modes to reasonably match the intrinsic absorption loss of the material.

The modal properties (density and radiative loss) of leaky modes can be engineered by control of the physical features, such as dimensionality, shape, of dielectric structures. [42, 43, 48] In the following, we will establish a quantitative understanding of the correlation between the physical features and the modal properties of leaky modes of dielectric structures. While in principle the physical feature may have an infinite amount of varieties, we focus on elucidating the correlation using a couple of representative structures. We will also evaluate the solar absorptions in the structures and demonstrate that the difference in the solar absorption can be correlated to the difference in the modal properties of leaky modes. The comparisons of solar absorptions are all made between structures with comparable geometrical cross-sections in the direction perpendicular to the incident solar light (referred as incident cross-section) or comparable volumes of absorbing materials.

1. Dimensionality: 1D vs. 0D structures

The main effect of dimensionality is on the density of leaky modes. The 1D and 0D structures here are defined to have subwavenlength dimensions in one and two directions in the plane perpendicular to the incident solar light, respectively. For instance, a horizontal nanowire is considered as a 1D structure, while a vertical nanorod may be better considered as a 0D structure. The number of leaky modes with resonant wavelength in the range of [$\lambda_1$, $\lambda_2$] can be written as $2\pi An^2(1/\lambda_1^2-1/\lambda_2^2)$ and $8\pi Vn^3(1/\lambda_1^3-1/\lambda_2^3)/3$ for 1D and 0D structures, respectively, where $A$ is the area of the 1D structure in the transverse direction and $V$ is the volume of the 0D structure.[45] These expressions may also be reasonably applied to heterostructures that combine absorbing and non-absorbing materials, for instance, core-shell structures, in which $A$ or $V$ would be the area or volume of the absorbing materials. This is because only the leaky modes associated with the absorbing part in heterostructures may substantially contribute to the solar absorption.

The effect of dimensionality can be best illustrated by examining the solar absorption. Figure 2a shows the calculated solar absorptions for single a-Si circular nanowires (1D) and spherical nanoparticles (0D) as a function of the radius $r$. To make sure that the solar absorptions are compared between the same volume of materials, the calculated solar absorption of the nanowire is multiplied with a length of $4r/3$, which comes from the volume relationship $\pi r^2 \times 4r/3 = 4\pi r^3/3$. We can find that the solar absorption of the 0D structure is generally larger than that of the 1D structure, except at small radii, such as < 50 nm. The different solar absorptions and associated dependence on the radius are rooted in the number of leaky modes. The number of leaky modes in 1D structures in the wavelength of interest [400 nm, 800 nm] ($2\pi An^2(1/\lambda_1^2-1/\lambda_2^2)$) is generally smaller than that in 0D structures ($8\pi Vn^3(1/\lambda_1^3-1/\lambda_2^3)/3$), but can turn to be larger when the radius is smaller than $9n\,(1/\lambda_1^2-1/\lambda_2^2)/[16(1/\lambda_1^3-1/\lambda_2^3)] \approx 50$ nm.

The effect of the different number of leaky modes can be further evidenced by the absorption spectra of 1D and 0D a-Si nanostructures with the same radius. For instance, a 100-nm radius 0D nanoparticle shows a substantial larger absorption efficiency $Q_{abs}$ than a 1D nanowire with the same radius (Fig. 2b). $Q_{abs}$ is defined as the ratio of the absorption cross-section $C_{abs}$ with respect to the incident cross-section $G$, which is $2r$ and $\pi r^2$ for the nanowire and the nanoparticle, respectively, $Q_{abs} = C_{abs}/G$. We find out all the leaky modes in the nanoparticle and the nanowire using the analytic method that we demonstrated previously,[42, 43] and plot the leaky modes along with the absorption spectra in Fig. 2b for the convenience of comparison. In the given spectral range of 400 – 1000 nm, 30 leaky modes can be found in the 0D nanoparticle, while only 19 leaky modes in the 1D nanowire.

We can conclude that, with comparable volume and incident cross-section, 0D structures can typically provide higher solar absorption than 1D structures due to a larger density of leaky modes. While the given results in this work are for spherical and circular structures, we confirm that this conclusion can generally apply to structures with other shapes like rectangle (Fig. 2c).

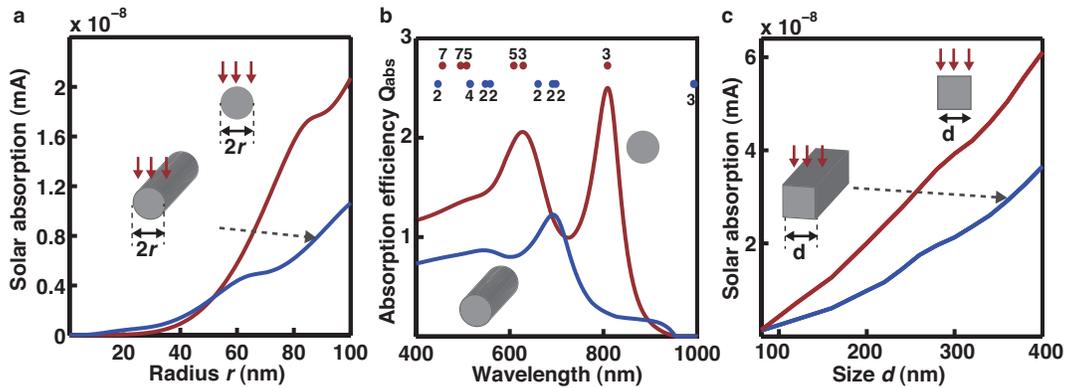

Figure 2. (a) Solar absorptions of 1D a-Si circular nanowires and 0D a-Si spherical nanoparticles as a function of the radius. (b) Spectral absorption efficiencies of the nanowire (blue curve) and the nanoparticle (red curve) in radius of 100 nm. Also plotted are the leaky modes in the nanowire (blue dots) and the nanoparticles (red dots). The given number indicates the degeneracy of each leaky mode. (c) Solar absorptions of 1D a-Si square nanowires and 0D a-Si cubic nanoparticles as a function of the size d.

2. Shape: Triangular, Rectangular, and Circular

We also examine the effect of shape on the modal properties of leaky modes and the solar absorption in dielectric nanostructures. We use 1D structures as an example, but the result can apply to 0D structures as well.

Fig. 3a shows the solar absorptions of 1D a-Si nanostructures with rectangular, circular, and triangular shapes. All these structures have the same incident cross-section as illustrated in Fig.3a inset. The heights of the rectangular and triangular structures are set to make them have the same volume as the circular one. These structures exhibit similar solar absorptions with small (<15%) difference, which is rooted in the comparable numbers and radiative losses of the leaky modes in these structures. As discussed in the preceding text, the density of leaky modes is dictated by the dimensionality. These 1D structures are expected to have a similar number of leaky modes given their identical dimensionality and material volume. Additionally, the radiative losses of leaky modes in these structures can be found comparable. Fig. 3b plots the radiative loss of typical leaky modes in these structures. To illustrate this notion in a general context, the radiative loss $q_{rad}$' is plotted as a function of the real part of the eigenvalue $N_{real}$, which is associated with the resonant wavelength $\lambda_0$ as $\lambda_0 = n\pi d/N_{real}$. Whereas an individual leaky mode could bear certain difference in the different shapes, the radiative losses of a large number of leaky modes, which is the typical case involved in solar absorption, show similar distributions regardless the shape of the structure. The comparable radiative losses of leaky modes can also be evidenced by the similar absorption efficiencies of these structures, as illustrated in Fig. 3c.

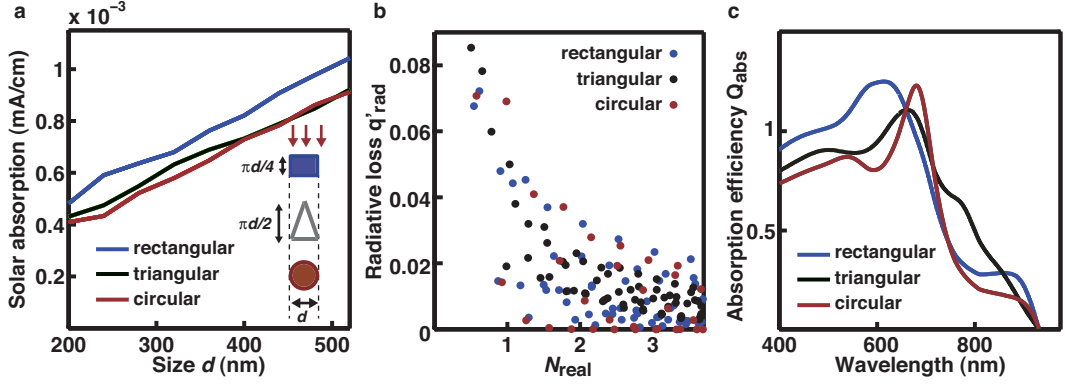

Figure 3. (a) Solar absorption of 1D a-Si structures with rectangular, triangular, and circular shapes as a function of the size $d$. All the structures have the same incident cross-section and the same volume of absorbing materials. (b) Modal properties of leaky modes in the rectangular, triangular, and circular structures. The horizontal and vertical axes are the real part of the eigenvalue and the radiative loss of leaky modes, respectively. (c) Spectral absorption efficiency of the rectangular, triangular, and circular structures with a size d of 200 nm.

We further examine the rectangular and triangular structures with different size ratios. Fig. 4a show the solar absorption of rectangular a-Si structures with the same volume (the area $a \times b$ is the same, where $a$ and $b$ are the height and the width of the structure, respectively) of materials but different size ratios $a:b$. Without losing generality, the size ratios of the rectangular structures given in Fig. 4a are set to be 1/3, 0.785, 1, and 3, respectively. Because different size ratios may give rise to different incident cross-section, for a fair comparison, we also normalize the calculated solar absorption with respect to the incident cross-section, which indicates the solar absorption per unit incident area (Fig. 4b). We can find that the structure with a larger incident cross-section has larger solar absorption per volume of materials (Fig. 4a) but smaller solar absorption per incident area (Fig. 4b). We have demonstrated that the radiative loss of leaky modes only show trivial change within the given size ratio.[44] The difference in the solar absorption can be mainly correlated to the coupling directivity of incident solar light with leaky modes.[49] A thorough discussion would be out of the scope of this work, but we can understand these different solar absorptions from an intuitive perspective. A larger incident cross-section may facilitate the coupling of incident solar light with leaky modes due to a larger exposed area to the incidence, which may subsequently give rise to larger solar absorption. Meanwhile, given the same amount of materials, a larger incident cross-section may result in a smaller height, which means a smaller number of leaky modes and smaller solar absorption per incident area. We can find a similar dependence of the solar absorption on the size ratio in triangular structures (Fig. 4c-d).

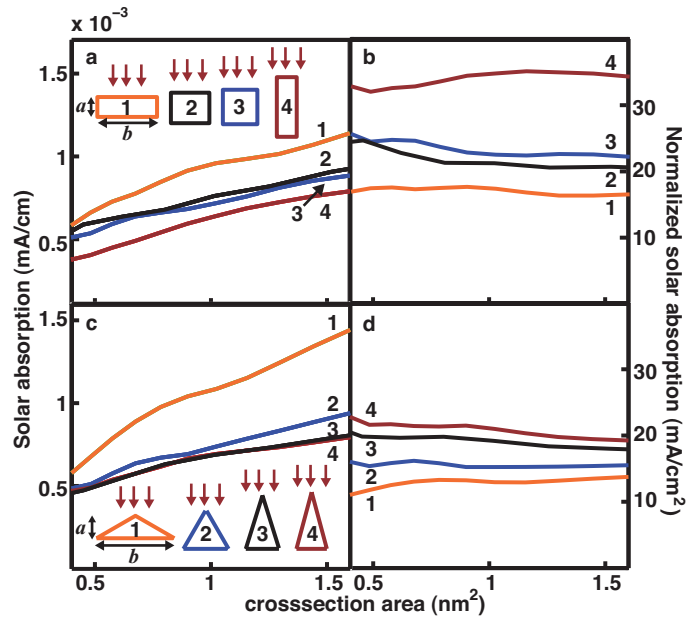

Figure 4. (a) Total solar absorption and (b) normalized solar absorption of 1D a-Si rectangular structures with different size ratios ($a/b$) of 1/3, 0.785, 1, and 3, which are labeled as structure 1, 2, 3, 4, respectively. All the structures have the same volume of absorbing materials. The inset is a schematic illustration of the structures and incident geometry. (c) Total solar absorption and (d) normalized solar absorption of 1D a-Si triangular structures with different size ratios ($a/b$) of 0.289, 0.866, 1.57, 1.87. All the structures have the same volume of absorbing materials. The inset is a schematic illustration of the structures and incident geometry.

We can reasonably conclude that the shape of dielectric structures does not substantially affect the modal properties of leaky modes from a statistics point of view. As a result, for a given volume of active materials, different shapes with comparable incident cross-sections may all have similar solar absorption. The shape with a larger incident cross-section may result in larger solar absorption per volume of materials but smaller solar absorption per incident area.

3. Heterostructures: core-shell and quasi-core-shell

Of our interest is the heterostructure that consists of absorbing and non-absorbing materials, such as a-Si and ZnO. The non-absorbing materials cannot absorb solar light by themselves, but may help enhance the solar absorption when forming heterostructures with absorbing materials. We have previously demonstrated that a-Si (core)-ZnO(shell) heterostructures can substantially boost the solar absorption in the a-Si core.[6] Other groups also report that heterostructures like void resonators, coated spheres/particles, may help enhance the solar absorption.[18-21, 31, 32] We now examine the improved absorption enhancement of heterostructures from the perspective of leaky modes.

We investigate two types of heterostructures, absorbing materials partially or fully coated by nonabsorbing materials (type A), and nonabsorbing materials partially or fully coated by absorbing materials (type B). Again, we use 1D structures as an example, but the results can apply to 0D structures as well.

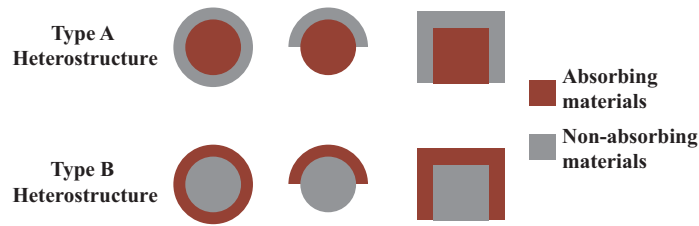

Figure 5. Schematic illustration for the heterostructures studied.

Fig. 6a shows the calculated solar absorption for a-Si structures coated by non-absorbing materials (type A) as a function of the size of the a-Si core. The solar absorption of a pure a-Si structure is also given as a reference. All the structures have the same volume of a-Si materials. Without losing generality, the non-absorbing coating is set to be in thickness of 70nm and with a constant refractive index of 2, which is close to that of ZnO. We can see that the solar absorptions of the fully and partially coated structures both are substantially larger than that of the pure a-Si structure. This improvement can be more clearly seen in absorption spectra. Fig.6b shows the spectral absorption efficiencies of these structures with a radius of 100 nm in the a-Si part, which confirms an improved absorption in a-Si materials with the presence of the full or partial coating. For visual convenience, the absorption spectrum of the rectangular heterostructure is not given.

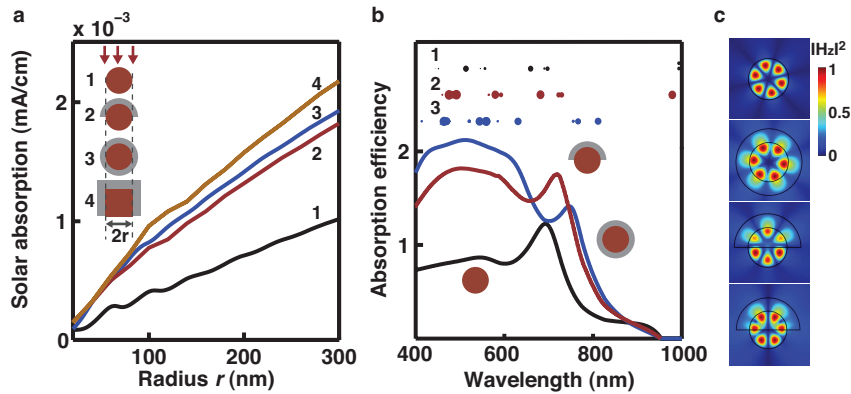

Figure 6. (a) Solar absorption of pure (structure 1, black curve), partially coated (structure 2, red curve), fully coated (structure 3, blue curve), and coated rectangular (structure 4, orange curve) a-Si structures as a function of the radius of the a-Si part. The inset schematically illustrates the structures and incident geometry. (b) Spectral absorption efficiencies of the circular structures (structure 1, 2, 3) with 100 nm in the radius of the a-Si core. Also plotted are the leaky modes of the structures, fully coated (blue dots), partially coated (red dots), and pure (black dots). The size of the dot is proportional to the radiative loss of the leaky mode. For visual convenience, the degeneracy of the leaky modes is not given. (c) The distribution of magnetic field intensity $|H_z|^2$ ($z$ is along the longitudinal direction of the 1D structure) of a typical mode $TE_{31}$ in the pure structure (top), fully coated (second top) and partially coated (the bottom two) structures.

We have previously demonstrated that the dielectric shell of an absorbing nanowire may essentially act as an anti-reflection coating.[6] The role of the dielectric shell can be more generally understood from the perspective of leaky modes. We calculate the leaky modes involved in the absorption of these structures using analytical or numerical (COMSOL) techniques.[42-44] We can find one-by-one correlation between the leaky modes in the coated structures and those in the pure a-Si structure, which is judged by the similarity in eigenfield distribution. Table 1 lists the resonant wavelength, eigvenvalue, and radiative loss of typical leaky modes in these structures. It is worthwhile to note that by coating a dielectric shell may actually create new leaky modes, but we only consider the leaky modes evolved from the original modes in the a-Si core, which typically show electromagnetic fields mainly distributed in the a-Si materials and are believed to contribute the majority of the absorption in the heterostructure.

While the coating may cause certain changes in the resonant wavelength of leaky modes (Table 1), the improved absorption enhancement mainly results from the increase in the radiative loss. All the leaky modes see increase in radiative loss after the coating, for instance, the radiative loss of the mode $TM_{21}$ (the nomenclature of leaky modes can be seen in the note of Table 1) changes from 0.014 in the pure structure to around 0.024 in the coated structures. To intuitively illustrate the correlation between the radiative loss of leaky modes and the absorption, we plot the leaky modes along with the absorption spectra with the size of each dot indicates the radiative loss (Fig. 6b). The increase in the radiative loss can be clearly seen from the eigenfield distribution of leaky modes. Fig. 6c show the distribution of the magnetic field intensity of a typical leaky mode $TE_{31}$ in the pure and coated a-Si structures. The field in the coated structure is obviously more spread than that in the pure structure. Correspondingly, the radiative loss of this mode can be found chaneing by one order of magnitude from 0.0019 in the pure structure to 0.016 or 0.025 and 0.017 in the half-coated and fully-coated structures, respectively.

Table 1. Leaky modes in pure, partially and fully coated a-Si nanostructures

| | Pure | | | Partially coated | | | Fully coated | | |
|---|---|---|---|---|---|---|---|---|---|
| | $\lambda_0$ (nm) | Eigenvalue | q | $\lambda_0$(nm) | Eigenvalue | $q'_{rad}$ | $\lambda_0$(nm) | Eigenvalue | $q'_{rad}$ |
| $TM_{21}$ | 692 | 3.71-0.05i | 0.014 | 724 | 3.49 - 0.079i | 0.023 | 753 | 3.31 - 0.084i | 0.025 |
| | | | | 724 | 3.49 - 0.083i | 0.024 | | | |
| $TM_{02}$ | 661 | 3.95-0.27i | 0.069 | 683 | 3.77 - 0.63i | 0.17 | 543 | 5.11 - 0.54Ii | 0.10 |
| $TM_{31}$ | 548 | 5.03-0.013i | 0.0027 | 567 | 4.82 - 0.044i | 0.009 | 584 | 4.63 - 0.044i | 0.01 |
| | | | | 566 | 4.83 - 0.036i | 0.008 | | | |
| $TM_{12}$ | 515 | 5.45-0.22i | 0.041 | 482 | 5.81 - 0.33i | 0.056 | 469 | 6.02 - 0.36i | 0.06 |
| | | | | 485 | 5.85 - 0.29i | 0.049 | | | |
| $TM_{41}$ | 446 | 6.30 - 0.003i | 0.0004 | 462 | 6.10 - 0.023i | 0.0038 | 477 | 5.92 - 0.015i | 0.003 |
| | | | | 462 | 6.1 - 0.014i | 0.002 | | | |
| $TE_{11}$ | 699 | 3.66-0.176i | 0.044 | 731 | 3.45 - 0.27i | 0.077 | 764 | 3.25 - 0.35i | 0.11 |
| | | | | 724 | 3.49 – 0.24i | 0.068 | | | |
| $TE_{21}$ | 558 | 4.92-0.063i | 0.012 | 594 | 4.52 - 0.17i | 0.039 | 639 | 4.21 - 0.26i | 0.061 |
| | | | | 585 | 4.61- 0.17i | 0.038 | | | |
| $TE_{02}$ | 515 | 5.45-0.223i | 0.042 | 492 | 5.73- 0.44i | 0.077 | 463 | 6.08 - 0.50i | 0.083 |
| $TE_{31}$ | 453 | 6.21-0.013i | 0.0019 | 489 | 5.76 - 0.089i | 0.016 | 584 | 5.38 - 0.091i | 0.017 |
| | | | | 483 | 5.84 - 0.14i | 0.025 | | | |
| $TE_{12}$ | N/A | 6.88 - 0.240i | 0.035 | 477 | 5.91 - 0.53i | 0.089 | N/A | 7.28 - 0.27i | 0.038 |
| | | | | N/A | 7.10 - 0.33i | 0.046 | | | |

Note: [1.] TE and TM refer to transverse electric and transverse magnetic polarizations, respectively. The two subscript numbers indicate the mode number and the order number of the leaky mode.[42, 43] [2.] Most of the given modes in the pure and the fully coated structures have a dual degeneracy except the $TM_{02}$ and $TE_{02}$ modes. But the dual degeneracy disappears in the partially coated structure.

Similar improvement in the absorption enhancement can be found in type B structures, and can also be correlated to increase in the radiative loss of leaky modes. Fig.7a shows the solar absorption of a full or partial a-Si coating on a non-absorbing core. Without losing generality, we set the radius of the core and the thickness of the shell in ratio of 4:1, and set the refractive index of the core to be 1.5, which is close to that of $SiO_2$. The rectangular structure is set to have similar dimensions to the circular ones as illustrated in Fig. 7a inset. The solar absorptions of the a-Si coatings, which have a much smaller volume of active materials, can be found very close to that of the pure a-Si structure. Fig. 7b shows the absorption spectra of the coatings and the pure a-Si structure. For simplicity, only the spectra of the circular structures are given. The a-Si coatings may have comparable or even larger absorption than that of the pure structure in certain wavelength ranges, for instance, from 580 nm to 620 nm (Fig. 7b). Again, this can be correlated to the increase of the radiative loss of leaky modes. Fig. 7c plots the leaky modes of these structures with resonant wavelength in the range of 580 - 620nm. The number of leaky modes in the pure structure (black dots) is apparently larger than both of the full (blue dots) and partial (red dots) coatings. However, there are multiple leaky modes in the a-Si coatings with radiative loss substantially larger than those in the pure structure. The gain from the increase of radiative loss may substantially offset the loss caused by the decrease in the number of leaky modes.

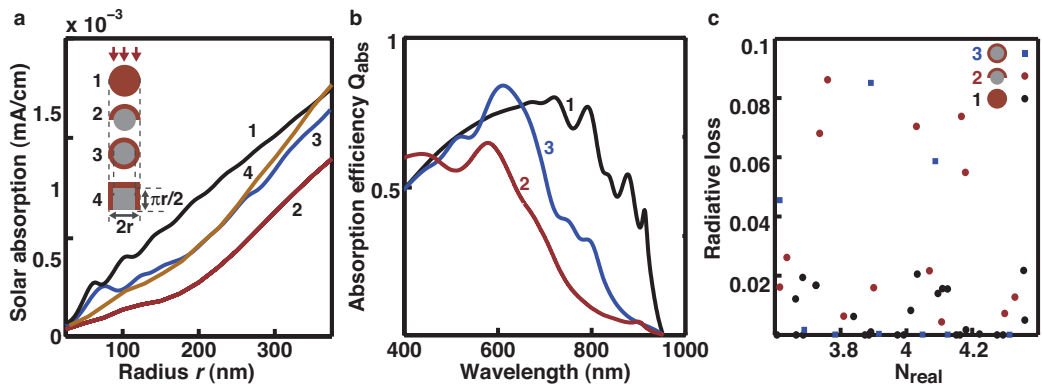

Figure 7. (a) Solar absorption of pure (structure 1, black curve), partial coating (structure 2, red curve), ful coating (structure 3, blue curve), and rectangular partial coating (structure 4, orange curve) a-Si structures as a function of the size $r$. The inset schematically illustrates the structures and incident geometry. The core and the thickness of the shell is set to be in a size ratio of 4:1. The refractive index of the core is 1.5. (b) Spectral absorption efficiencies of the circular structures (structure 1, 2, and 3) with $r = 375$ nm. (c) The modal properties of leaky modes in the structures with resonant wavelengths in the range of 580 - 620 nm, full coating (blue dots), partial coating (red dots), and pure (black dots). For visual convenience, the degeneracy of the leaky modes is not given. Generally, most of the leaky modes in the pure structure and the full coating have a degeneracy of two, but the mode in the partial coating does not have the degeneracy.

Our analysis indicates that heterostructuring poses a promising strategy for the engineering of the radiative loss of leaky modes. While the non-absorbing materials cannot absorb solar light by themselves, heterostructuring them, which typically have a lower refractive index, with absorbing materials can change the refractive index contrast of the absorbing materials and hence the radiative loss of leaky modes. The effect on the radiative loss of leaky modes can essentially accounts for all the improvements in the solar absorption of heterostructures reported in literature.[6, 18-21, 31]

4. Single nanostructures and an array of nanostructures

Of the most interest for practical applications is a large scale array of nanostructures. As the design of single nanostructure absorbers would be much less time-consuming, understanding the correlation between the solar absorption of single nanostructures and that of a nanostructure array can significantly facilitate the rational design of solar absorbers for practical applications.

We use single 1D structures and a periodic array of 1D structures as an example to address this issue from the perspective of leaky modes. We find that, with a reasonable inter-nanostructure spacing (typically not too small compared to the size of each individual nanostructure), the leaky modes in the array are similar to those in the single nanostructure. Fig. 8 shows the absorption spectra of single a-Si 1D structures in radius of 100nm and an array of the nanostructure with a period of 400 nm (Fig.8 inset). For a fair comparison, the absorption efficiency of the array is calculated with respect to the entire space including the inter-nanostructure opening, and that of the single nanostructure is calculated by normalizing its absorption cross-section with respect to a size of 400 nm, which is equal to the period of the array. We can find that the absorption spectra of the two structures are reasonably similar.

To understand the similarity, we calculate the leaky modes of the single nanostructure and the array, and list the calculated modal properties in Fig. 8. Most of the leaky modes in the nanostructure array show a one-by-one correlation with the mode in the single nanostructure, which is judged by a similar eigenfield distribution. Only one leaky mode in the nanostructure array cannot be correlated to any of the modes in the single structure, which is listed at the end of the table. The new mode can be recognized resulting from the Bragg scattering in the nanostructure array. This one-by-one correlation indicates that the majority of the leaky modes in the nanostructure array originate from the modes of single nanostructures. We can also find that the modes in both structures show similar radiative losses and resonant wavelengths (Fig. 8). As a result, each individual nanostructure in the array may absorb solar radiation in a similar way as the single nanostructure. This conclusion holds for any nanostructure array with interspacing not too small comparable to the size of individual nanostructures. Therefore, the solar absorption of single nanostructures can be used as a reference for the rational design of large-scale nanostructure array for applications in practical solar cells.

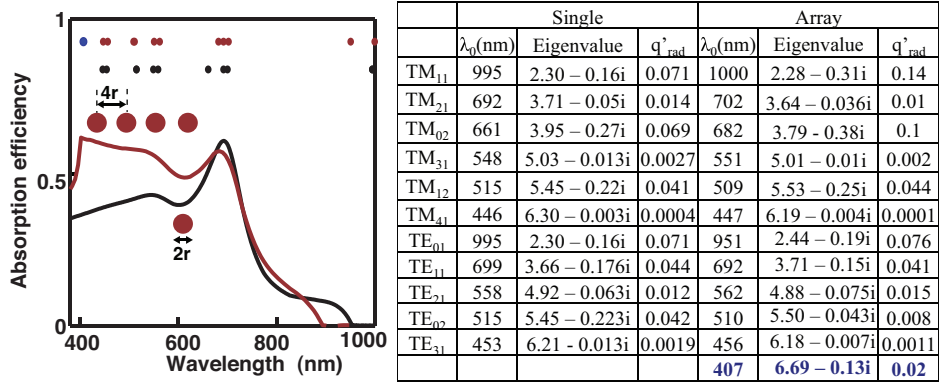

|  | Single | | | Array | | |
| --- | --- | --- | --- | --- | --- | --- |
|  | $\lambda_0$(nm) | Eigenvalue | $q'_{rad}$ | $\lambda_0$(nm) | Eigenvalue | $q'_{rad}$ |
| $TM_{11}$ | 995 | 2.30 – 0.16i | 0.071 | 1000 | 2.28 – 0.31i | 0.14 |
| $TM_{21}$ | 692 | 3.71 – 0.05i | 0.014 | 702 | 3.64 – 0.036i | 0.01 |
| $TM_{02}$ | 661 | 3.95 – 0.27i | 0.069 | 682 | 3.79 - 0.38i | 0.1 |
| $TM_{31}$ | 548 | 5.03 – 0.013i | 0.0027 | 551 | 5.01 – 0.01i | 0.002 |
| $TM_{12}$ | 515 | 5.45 – 0.22i | 0.041 | 509 | 5.53 – 0.25i | 0.044 |
| $TM_{41}$ | 446 | 6.30 – 0.003i | 0.0004 | 447 | 6.19 – 0.004i | 0.0001 |
| $TE_{01}$ | 995 | 2.30 – 0.16i | 0.071 | 951 | 2.44 – 0.19i | 0.076 |
| $TE_{11}$ | 699 | 3.66 – 0.176i | 0.044 | 692 | 3.71 – 0.15i | 0.041 |
| $TE_{21}$ | 558 | 4.92 – 0.063i | 0.012 | 562 | 4.88 – 0.075i | 0.015 |
| $TE_{02}$ | 515 | 5.45 – 0.223i | 0.042 | 510 | 5.50 – 0.043i | 0.008 |
| $TE_{31}$ | 453 | 6.21 - 0.013i | 0.0019 | 456 | 6.18 – 0.007i | 0.0011 |
|  |  |  |  | 407 | 6.69 – 0.13i | 0.02 |

Figure 8. (Left) Spectral absorption efficiencies of single a-Si 1D nanostructures with a radius of 100 nm and of an array of the 1D nanostructure with 400nm in the interspacing. Also plotted are the leaky modes in the single nanostructure (black dots) and the nanostructure array (red dots). The blue dot indicates the leaky mode in the nanostructure array that cannot be correlated to any of the modes in the single structure. (Right) tabulated modal properties for the leaky modes in the single nanostructure and the nanostructure array. Again, the one in blue indicates the leaky mode in the nanostructure array that cannot be correlated to any of the modes in the single structure.

As the final note, leaky mode engineering represents a general principle for the rational design of dielectric optical antennas with optimal solar absorption enhancement. Results of this work points out the design guideline, a) using 0D structures; b) the shape does not matter much; c) heterostructuring with non-absorbing materials is a promising strategy; d) the design of a large-scale nanostructure array can use the solar absorption of single nanostructures as a reference. It is worthwhile to emphasize again that key to maximize the solar absorption of a material is to engineer the radiative loss of leaky modes to reasonably match the instrinsic absorption loss of the material. More studies would be necessary to find out heterostructures that can provide leaky modes satisfying this requirement.

## Acknowledgements

This work is supported by start-up fund from North Carolina State University. L. Cao acknowledges a Ralph E. Power Junior Faculty Enhancement Award from Oak Ridge Associated Universities.